\documentclass[prl,twocolumn,showpacs,floatfix]{revtex4-1}
\usepackage[T1]{fontenc}
\usepackage[cp1250]{inputenc}
\usepackage{psfrag}
\usepackage{color}
\usepackage{graphicx}
\usepackage[english]{babel}
\newcommand{\bq}{\begin{equation}}
\newcommand{\eq}{\end{equation}}
\newcommand{\ba}{\begin{eqnarray}}
\newcommand{\ea}{\end{eqnarray}}
\newcommand{\D}{{\rm d}}

\begin{document}
\title{Diffusion with Stochastic Resetting}
\author{Martin R. Evans$^{(1,2)}$ and Satya N. Majumdar$^{(2)}$}
\affiliation{$^{(1)}$ SUPA, School of Physics and Astronomy, University of Edinburgh, Mayfield Road, Edinburgh EH9 3JZ, United Kingdom\\
$^{(2)}$ Univ. Paris-Sud, CNRS, LPTMS, UMR 8626, Orsay F-01405, France}
\noindent
\begin{abstract}
We study simple diffusion where a particle stochastically resets to its 
initial position at a constant rate $r$. A finite resetting rate leads
to a nonequilibrium stationary state with non-Gaussian fluctuations
for the particle position. We also show that the mean time to find
a stationary target by a diffusive searcher is finite and has a minimum 
value at an optimal
resetting rate $r^*$. Resetting also 
alters fundamentally the late time decay of the survival 
probability of
a stationary target when there are multiple searchers: while the {\em 
typical} survival probability decays exponentially with time, the {\em 
average} decays as a power law with an exponent depending continuously
on the density of searchers.   
\end{abstract}
\maketitle

`Stochastic resetting' is a rather common process in everyday life.
Consider searching for some target such as, for example, a face in a 
crowd or one's misplaced keys at home.
A natural tendency is, on having searched unsuccessfully for a while,
to return to the starting point 
and recommence the search. In this Letter we explore
the consequences of such resetting on perhaps the most simple and 
common 
process in nature, namely, the diffusion of a single or a multiparticle system.
We show that a nonzero rate of resetting has a rather 
rich and dramatic effect on the diffusion process.

The first major effect of resetting shows up in the position distribution
of the diffusing particle. In the absence of resetting, it has the usual   
Gaussian distribution whose width grows diffusively
$\sim \sqrt{t}$ with time. Upon switching on 
a nonzero resetting rate $r$ to its initial position, this
time-dependent Gaussian distribution gives way to a globally 
current-carrying {\em nonequilibrium stationary state} (NESS) with {\em 
non-Gaussian} fluctuations, given in Eq. (\ref{pss}). The process of resetting manifestly violates detailed 
balance and thus provides an appealingly simple example  of a NESS.

Resetting also has a profound consequence 
on the first-passage properties of a diffusing particle. 
The study of first-passage problems and survival probabilities of
diffusing particles arises in diverse subjects such as in 
reaction-diffusion kinetics, predator-prey dynamics~\cite{Redner},
as well as in persistence in nonequilibrium systems~\cite{pers-review}.
Such problems are
fundamental to nonequilibrium statistical mechanics as they
involve irreversible processes not obeying 
detailed balance. 
Related models  are also relevant to the study of search strategies
in ecology or sampling techniques for the characterisation of complex
networks.  For example,  intermittent 
searches involve diffusive motion combined with long range movements
of the searcher and mimic the scan and relocation phases
of foraging animals~\cite{Bell,BCMSV05,LKMK08,LBMV09,BC09,GLWB09}.

A well-studied problem is the mean time for a stationary target at the origin 
to be absorbed by a single diffusing particle (trap) or a team of 
diffusing traps
distributed with uniform density.  Many significant results,
such as the fact that the mean time to find the target by a single diffusing
particle diverges and that the survival probabilty of the target
decays with time as a stretched exponential in the presence of a
finite density of diffusing particles, are well-established 
when there is no resetting~\cite{Redner,Klafter,bb}.
\begin{figure}
\includegraphics[width=.7\hsize]{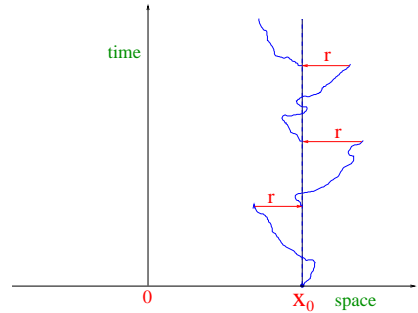}
\caption{Schematic space-time trajectory of a one dimensional Brownian motion 
that starts at $x_0$ and resets stochastically to its initial position 
$x_0$ at rate $r$.}
\label{fig1:reset}
\end{figure}

Our results,
summarised in this paragraph, 
show that these statistical  properties of the search 
for a stationary target  
by diffusing traps (searchers)
are fundamentally altered
when the searchers
reset  stochastically  to their initial positions.
In the case of 
a single diffusing trap,  the mean time to find the target
at the origin, given by Eq. (\ref{T1d}),  becomes {\em finite} in the presence of a non-zero resetting rate 
$r$ 
and as a function of $r$, has a minimum at a nontrivial value $r^*$
given by Eq. (\ref{z*}).
Thus there is an optimal resetting rate that makes the target search
most efficient.  The survival probability of
the target decays exponentially with nontrivial decay rate 
summarized in Eqs. (\ref{Qzll},\ref{Qzgg}).
For the case of mutiple traps distributed with uniform
density $\rho$ and each diffusing independently with a diffusion constant 
$D$, 
the effect of resetting is even more dramatic.
For example, in one dimension and in the absence of resetting, both the {\em 
typical} and the {\em average}
survival probability of the target decay with time $t$ as a stretched 
exponential, $\sim \exp (-\lambda \rho \sqrt{D\,t})$, 
albeit with two different values
of the decay constant $\lambda$~\cite{Klafter,bb}. In contrast, 
we find that when one 
switches on 
the resetting  the typical survival probability of the target 
asymptoticaly decays
faster with time,  as an  exponential, Eq. (\ref{Psq}).
On the other hand, the average survival probability 
decays more slowly  asymptotically, as a power law with a nonuniversal exponent 
that varies continuously with density $\rho$,  Eq. (\ref{Psa}). This slower algebraic decay
can be attributed to contributions arising from rare, extreme events.

We now present a derivation of our results.
For simplicity, we focus below on the one-dimensional case, though 
generalization to higher dimensions is straightforward and will
be mentioned at the end.
To begin with we consider the simple case of a single particle
diffusing in one dimension with diffusion constant $D$ and stochastically 
resetting to its initial
position at a constant rate $r$ (see Fig. 
\ref{fig1:reset}).
The Master equation for $p(x,t|x_0)$, the probability that the particle
is at $x$ at time $t$, having begun from $x_0$ at time 0, simply reads
\begin{equation}
\frac{\partial p(x,t|x_0)}{\partial t}
= D\frac{\partial^2 p(x,t |x_0)}{\partial x^2}
 - r p(x,t|x_0)+ r\delta(x-x_0)
\label{me}
\end{equation}
with initial condition $p(x,0) = \delta(x-x_0)$.
The second and third terms on the rhs 
and represent a negative
probability flux  $r p(x,t|x_0)$  out of each point $x$ 
and a corresponding  positive probability flux into $x_0$ which sums to $r$. 
It is easy to show that in the long-time limit the particle attains a 
stationary distribution 
\begin{equation}
p_{\rm st}(x|x_0) = \frac{\alpha_0}{2} \exp( - \alpha_0 |x-x_0|)
\label{pss}
\end{equation}
where  $\alpha_0 = \sqrt{r/D}$
is an inverse length scale corresponding
to the typical distance diffused by the particle between resets.
The distribution (\ref{pss}) is evidently non-Gaussian, with a cusp
at $x=x_0$. Since resetting creates a source of probabality at
$x_0$ while probability is lost from all $x\neq x_0$, there is
circulation of probability even at long times, making 
(\ref{pss}) a non-equilibrium stationary state.

The result (\ref{pss}) has a  simple and appealing  interpretation in terms 
of a
renewal process. The time evolution of the diffusive particle
(see Fig. (\ref{fig1:reset})) is indeed a renewal process, where
each resetting event to $x_0$ renews the process at a Poisson rate $r$ and 
between two consecutive
renewal events, the particle undergoes free diffusion. Consider
the particle at some fixed time $t$. If $\tau$ denotes the time
elapsed since the last renewal, the position distribution is
simply Gaussian, $e^{-(x-x_0)^2/{4D\tau}}/\sqrt{4\pi D\tau}$.
Clearly $0\le \tau\le t$ is a random variable and for large $t$,
$\tau$ is simply exponentially distributed, ${\rm Prob}(\tau)= re^{-r\tau}$.
Thus, for large $t$, the stationary position distribution of the
particle is obtained by averaging over $\tau$
\begin{equation}
p_{\rm st}(x|x_0)= \int_0^{\infty}d\tau\, r\,e^{-r\,\tau}\, 
\frac{e^{-(x-x_0)^2/{4D\tau}}}{\sqrt{4\pi D\tau}}
\label{renewal1}
\end{equation}
which precisely yields the result in (\ref{pss}). 

We now turn to the first-passage properties through the origin of the
diffusing particle in the presence of resetting. 
The origin can be thought of as a
stationary target which is absorbed when the diffusing particle, starting
initially at $x_0>0$ at $t=0$, hits the origin for the first time.
The diffusing particle also resets to its initial position $x_0$ at a 
constant rate $r$. To solve the first-passage problem, it 
is advantageous to use the backward Master equation approach where one 
treats the {\em initial}  position as a variable. Hence, we first set the initial
position to be at $x>0$, different from the resetting position $x_0$,
solve the problem with arbitrary $x$ and $x_0$ and eventually set 
$x=x_0$. Let $Q(x,t)$ denote the survival probability or persistence of the
target up to time $t$ (it also depends implicitly on the resetting 
position $x_0$). This is 
then also the probability that the
diffusing particle, starting at $x$, has not hit the origin up to time $t$.
The backward Master equation (where the initial position
$x$ is the variable)
then reads
\begin{equation}
\frac{\partial Q(x,t)}{\partial t}
= D\frac{\partial^2 Q(x,t)}{\partial x^2}
 - r Q(x,t) + rQ(x_0,t)
\label{bme}
\end{equation}
with boundary and initial conditions $Q(0,t)=0$, $Q(x,0)=1$. 
The second and third terms on the rhs correspond 
to the resetting of the {\em initial} position 
from $x$ to $x_0$,
which implies
a loss of probability from $Q(x,t)$ and a gain of probability
to $Q(x_0,t)$.
The Laplace transform $q(x,s) = \int_0^\infty\D t\ {\rm e}^{-st} Q(x,t)$
satisfies
\begin{equation}
D\frac{\partial^2 q(x,s)}{\partial x^2}
 - (r+s) q(x,s) = -1 - rq(x_0,s)
\label{bmelt}
\end{equation}
which can be exactly solved by noting that the general solution
is
$q(x,s) = A {\rm e}^{\alpha x} + B {\rm e}^{-\alpha x} 
+(1 + rq(x_0,s))/(r+s)$
with $\alpha = \sqrt{(r+s)/D}$.
Now $q(x,s)$ remains finite as $x \to \infty$ therefore $A=0$. 
The condition $q(0,s)=0$ fixes the constant 
 $B = -(1 + rq(x_0,s))/(r+s)$ 
and $q(x_0,s)$ is determined self-consistently as
\begin{equation}
q(x_0,s) = \frac{1 -
\exp(-\alpha x_0)}{s+ r \exp(-\alpha x_0)}\;.
\label{qlts}
\end{equation}

The mean first-passage time, $T(x_0)$, may be obtained from
$T(x_0)  = - \int_0^\infty \D t\, t \frac{\partial Q(x_0,t)}{\partial t}=
q(x_0,s=0)$
which yields
\begin{equation}
T(x_0) = \frac{1}{r} \left(\exp(\alpha_0 x_0) -1\right)
\label{T1d}
\end{equation}
where we recall $\alpha_0=\sqrt{r/D}$.
The first point to note is that $T$ is finite for $r>0$ and diverges
as $r \to 0$ as $T \sim r^{-1/2}$, which recovers the well-known result
that the mean time for a diffusive particle to reach the origin (in the
absence of resetting) is infinite.
Also $T$ diverges as $r \to \infty$, the explanation
being that as the reset rate increases the diffusing particle 
has less time between resets to reach the origin.

Now consider $T$ as a function of $r$ for a fixed $x_0$.
Since  $T$ diverges as $r \to 0$ and $r\to \infty$
it is clear that there must be a minimum of $T$ with
respect to $r$ at an optimal value $r^*$.
We consider the 
dimensionless variable $z = \alpha_0 x_0$,
the ratio of the distance of the intial site from the target
to the length diffused between resets, with which our results may be
simply expressed.
Then the condition for the minimum, $\D T/\D r=0$, reduces to
the transcendental equation 
\begin{equation}
\frac{z^*}{2} = 1 - {\rm e}^{-z^*}
\label{z*}
\end{equation}
which has a unique non-zero
solution $z^*= 1.59362....$. Thus, there is an optimal resetting
rate $r^*= (z^*)^2D/x_0^2$ in one dimension that minimizes the
search time to find the target at the origin.

It is  difficult to invert the
Laplace transform (\ref{qlts}) explicitly for all parameters. 
However, 
one can deduce the large $t$ asymptotic behaviour for fixed
parameters $r,x_0$. Generally this is determined by the singularity,
in the complex $s$ plane, of 
$q(x_0,s)$ with largest real part.
From (\ref{qlts}) there  will be a pole in $q(x_0,s)$
when
\begin{equation}
s+ r {\rm e}^{-\alpha x_0} =0
\label{s0}
\end{equation}
 which may be rewritten as
$s= r(u-1)$
where $u$ satisfies
\begin{equation}
u = 1- \exp(- u^{1/2} z)\;.
\label{ueq}
\end{equation}
An addition to the trivial solution $u=0$ (which corresponds to 
the branch point at $s=-r$) there is 
exactly one non-trivial solution $u_0$ of (\ref{ueq})
where $0 < u_0 <1$. Thus 
$q(x_0,s)$ has a simple pole at 
$s_0= -r(1-u_0)$ the residue of which determines the asymptotic behaviour as
\begin{eqnarray}
Q(x_0,t) &\simeq& 
{\rm e}^{s_0t} \frac{2 u_0^{3/2}}{ z u_0 + 2 u_0^{1/2} -z}\;.
\label{Qlt}
\end{eqnarray}
Different limiting values of $u_0$ and hence $s_0$ can be deduced.
For $z \ll 1$, $u_0 \simeq z^2$ and $s_0 \simeq -r + rz^2$
whereas for $z \gg 1$, $u_0 \simeq 1- {\rm e}^{-z}$ and 
$s_0 \simeq -r{\rm e}^{-z}$.
These results imply that 
the survival probability decays asymptotically as
\begin{eqnarray}
Q(x_0,t) &\simeq& 2 z^2 \exp( -rt(1-z^2+...))\;\;\mbox{for}\;\; z\ll 1
\label{Qzll}\\
Q(x_0,t) &\simeq& \exp( -rt \exp(- z))\;\;\mbox{for}\;\;z \gg 1\;.
\label{Qzgg}
\end{eqnarray}
One may relate the large $z$ result to the Gumbel distribution
for the extremum of independent random variables~\cite{Gumbel}
as follows. In time $t$ the mean number of resets to position $x_0$ is
$N=rt$. For the target to survive requires that between each reset the 
maximum
distance to the left attained by the particle is less than $x_0$.
In a given reset interval of length $\tau$, the particle undergoes free
diffusion around $x_0$. Hence, the probability that the minimum
is bigger than $m$ is ${\rm 
erf}\left(\frac{|m-x_0|}{\sqrt{4D\tau}}\right)$~\cite{Redner}
and the pdf of the minimum $m$ is ${\rm 
Prob}(m)=e^{-(m-x_0)^2/{4D\tau}}/\sqrt{\pi D\tau}$ with $m\le x_0$. Next, we 
average over
$\tau$ drawn from the exponential distribution ${\rm Prob}(\tau)=r 
e^{-r\tau}$.
Thus, the effective pdf of the maximum over each reset interval is
then $p(m)= \alpha_0 e^{-\alpha_0(x_0-m)}$ where $m\le x_0$.
Taking the $N$ intervals as statistically independent, the probability
that the minimum of all $N$ intervals stays above $0$ is
$[\int_0^{x_0} p(m) dm]^N$. Performing the integral
and taking large $z$ limit yields
$\approx \exp\left[-rt 
\exp(-z)\right]$.  

We now turn to 
the survival probability of a stationary target at the origin
in the presence of many independent searchers (diffusive particles).
We consider $N$ diffusive particles $i=1,\ldots, N$, each of which 
is reset independently to its initial position $x_i$ with  rate $r$.
The survival probability of the target is given by
\begin{equation}
P_s(t) = \prod_{i=1}^N Q(x_i,t)
\label{Psdef}
\end{equation}
where 
$Q(x_i,t)$ is the survival probability in the single searcher problem.

The initial positions $x_i$'s are assumed to be independent and each
distributed uniformly over the box $[-L/2,L/2]$. Consequently, $P_s(t)$
is a random variable. Its average is simply $P_s^{\rm av}(t)=\langle 
P_s(t)\rangle_x$ where $\langle \rangle_x$ denotes averages over $x_i$'s.
However, $P_s(t)$ for a {\em typical} initial configuration
is not captured by the average. The typical $P_s(t)$ can be extracted
by first averaging over the logarithm of $P_s(t)$ followed by 
exponentiating: $P_s^{\rm typ}(t)= \exp\left[\langle \ln P_s(t) \rangle_x 
\right]$. One can draw an analogy to a disordered system with $P_s(t)$
playing the role of partition function $Z$ and $x_i$'s as disorder variables.
Thus the average and typical behavior correspond respectively 
to the {\em annealed} (where one averages the partition function $Z$)
and the {\em quenched} (where one averages the free energy $\ln Z$)
averages in disordered systems.

In the annealed case we get
\begin{equation}
P_s^{\rm av}(t) = \langle  Q(x,t)\rangle_x^N = \exp N \ln[1- \langle 1- 
Q\rangle_x]\;
\end{equation}
where
\begin{equation}
\langle 1- Q\rangle_x = \frac{1}{L} \int_{-L/2}^{L/2}
\D x [1- Q(x,t)]\; .
\end{equation}
Letting $N,L \to \infty$
but keeping the density of walkers $\rho=N/{L}$ fixed, and using
the symmetry $Q(x,t)=Q(-x,t)$,
we obtain
\begin{equation}
P_s^{\rm av}(t) \to  \exp -2\rho \int_0^\infty\D x  [1-  Q(x,t)]
\equiv\exp -2\rho M(t)\;.
\label{Psann}
\end{equation}
The Laplace transform ${\tilde M}(s)=\int_0^{\infty} M(t) e^{-st} dt$
can be determined using (\ref{qlts})
\begin{equation}
{\tilde M}(s) =
\frac{r+s}{s r \alpha} \ln \left( \frac{s+r}{s}\right)\;.
\end{equation}
It may then be inverted~\cite{details} to obtain
\begin{equation}
M(t) = \left( \frac{D}{r}\right)^{1/2} \mu(y)\;,
\end{equation}
with $y=rt$, where
\begin{equation}
\mu(y) = \int_0^y \D v \frac{1- {\rm e}^{-v}}{v}
\left[ \mbox{erf}\left[ (y-v)^{1/2}\right] + \frac{{\rm e}^{-(y-v)}}{\sqrt{\pi(y-v)}}\right]\;.
\end{equation}
The  function $\mu(y)$ behaves  asymptotically as
\begin{eqnarray}
\mu(y) &\simeq& \frac{2 y^{1/2}}{\sqrt{\pi}}\qquad\mbox{for small}\quad y\\
& \simeq & \ln y + \gamma \qquad \mbox{for large}\quad y
\end{eqnarray}
where $\gamma$ is Euler's constant.
Thus  the long-time behaviour ($rt \gg 1$) of
the annealed survival probability is a power law with an exponent that
varies  continuously with the density
(\ref{Psann}) as
\begin{equation}
P_s^{\rm av}(t) \simeq A  t^{-2\rho (D/r)^{1/2}}\;,
\label{Psa}
\end{equation}
where $A$ is a constant that may be determined.
The short-time behaviour ($rt \ll 1$) is stretched exponential decay
$P_s^{\rm av}(t) \simeq \exp( -4\rho (Dt)^{1/2}/\sqrt{\pi} )$
as in the case of diffusion without resetting~\cite{Klafter,bb}.

In contrast, the typical behavior (the quenched case) 
$P_s^{\rm typ}(t)  = \exp \left[\langle \ln P_s(t) \rangle_x\right]$
can be expressed as
\begin{equation}
P_s^{\rm typ}(t) =   \exp  \sum_{i=1}^N  \langle \ln Q(x_i,t) \rangle_x
= \exp \left[   2\rho \int_0^{L/2} \D x \ln  Q(x,t)\right].  
\end{equation}
In the long time limit, we have using 
(\ref{Qlt})
\begin{equation}
\int_0^\infty \D x_0 \ln Q(x_0,t) \simeq
\mbox{Constant}
 \left[- t \int_0^\infty \D x_0 |s_0(x_0)|\right].
\end{equation}
The integral can be done in closed form~\cite{details} and we find
$\int_0^\infty \D x_0 s_0(x_0)=- (D r)^{1/2} 4(1-\ln 2)$.
Thus the  asymptotic decay of the quenched total survival probability 
is exponential
\begin{equation}
P^q_s(t) \sim \exp \left( - t \rho (D r)^{1/2} 8(1- \ln 2) \right).
\label{Psq}
\end{equation}
The correction to the argument of the exponential in
(\ref{Psq}) will come from the branch point at $s= -r$ in (\ref{qlts})
and is expected to give a contribution $O(t^{1/2})$.
The fact that the average and typical survival probabilities 
have distinct asymptotic behaviours 
reflects the strong  dependence on the initial conditions
whose memory is retained through resetting.

Our results can straightforwardly be extended to higher dimensions.
For example, in $d$-dimensions,
the second derivative term in the Master equation (\ref{me}) is
replaced by a $d$-dimensional Laplacian. The stationary solution in
$d$-dimensions is easy to find~\cite{details} 
\begin{equation} 
p_{\rm st}({\vec x}|{\vec x_0})= \frac{(\alpha_0)^d}{(2\pi)^{d/2}} 
(\alpha_0 |\vec x-\vec x_0|)^\nu K_\nu(\alpha_0 |\vec x-\vec x_0|)
\label{pssd}  
\end{equation}
where $\nu = 1- d/2$ and $K_\nu$ is the modified Bessel function.
Similarly, the mean absorption time
for an absorbing ball of radius $\epsilon$ at the origin is~\cite{details}
\begin{equation}
T(\vec x_0) = \frac{ 1}{r}\left[
\left( \frac{\epsilon}{|\vec x_0|}\right)^\nu 
\frac{K_\nu( \alpha_0 \epsilon)}
{ K_\nu( \alpha_0 |\vec x_0 |)} -1 \right].
\label{Td}
\end{equation}
The survival probability in case of multiple searchers, both annealed and 
quenched, can also be computed
in higher dimensions~\cite{details}.

In conclusion, in this Letter we have shown that introduction of
resetting events leads to rather rich effects on simple diffusion process.
In particular,  the effect of resetting on the efficient search   
for a stationary target both by a single or a team of searchers is
rather profound.
Our results can be extended in several directions, in particular
one may consider resetting rates $r(x)$ that depend on the position of the
particle~\cite{details}. It would also be interesting to study the
effect of resetting on other available search strategies based 
on nondiffusive processes such as L\'evy flights~\cite{BCMSV05,LKMK08}.

\begin{acknowledgements}
We thank R.~A. Blythe, B. Derrida and S. Janson for useful
discussions. 
\end{acknowledgements}

\end{document}